\documentclass[12pt,reqno]{amsart}                  

\usepackage[dvips]{graphicx}                        
\usepackage{latexsym}                               
\usepackage{amssymb}                                
\usepackage[colorlinks=true, hypertex, linkcolor=blue, urlcolor=blue, citecolor=blue]{hyperref}

\newcounter{maintheoremnumber}
\newtheorem{theorem}{Theorem}[section]                
\newtheorem{mainthm}[maintheoremnumber]{Theorem}
\newtheorem{lemma}{Lemma}[section]

\newtheorem{proposition}[lemma]{Proposition}

\renewcommand{\epsilon}{\varepsilon}

\newcommand{\bth}{\begin{theorem}}
\newcommand{\n}{\noindent}
\newcommand{\dis}{\displaystyle}

\newcommand{\im}{\, \textup{Im} \,}
\newcommand{\R}{\mathbb{R}}
\newcommand{\C}{\mathbb{C}}
\newcommand{\ba}{\begin{align*}}
\newcommand{\ea}{\end{align*}}

\begin{document}

\title{Density of Complex Critical Points of a Real Random $SO(m+1)$ Polynomial}
\author{Brian Macdonald}
\email{bmac@jhu.edu}
\thanks{Thanks to Bernard Shiffman, for his guidance, teaching, patience, and
for countless helpful meetings, conversations and suggestions.}

\maketitle

\begin{abstract}
    We study the density of complex critical points of a real random $SO(m+1)$ polynomial in $m$ variables.  In a previous paper \cite{bmzeros}, the author used the Poincar\'{e}-Lelong formula to show that the density of complex zeros of a system of these real random polynomials rapidly approaches the density of complex zeros of a system of the corresponding complex random polynomials, the $SU(m+1)$ polynomials.  In this paper, we use the Kac-Rice formula to prove an analogous result: the density of complex critical points of one of these real random polynomials rapidly approaches the density of complex critical points of the corresponding complex random polynomial.  In one variable, we give an exact formula and a scaling limit formula for the density of critical points of the real random $SO(2)$ polynomial as well as for the density of critical points of the corresponding complex random $SU(2)$ polynomial.

\end{abstract}

\section{Introduction}

\begin{figure}
            \centering
                \includegraphics[width=0.4\textwidth]{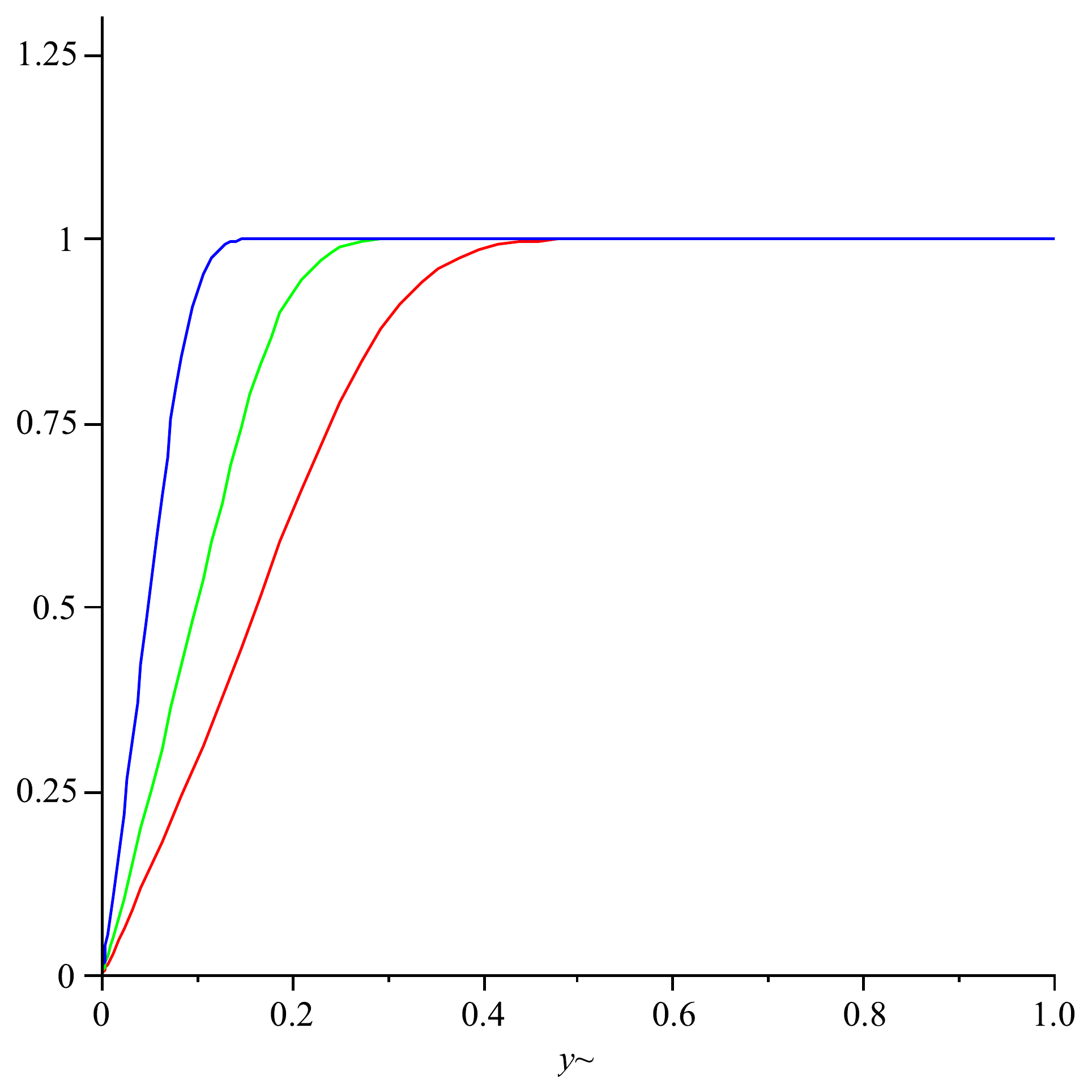}
                \includegraphics[width=.4\textwidth]{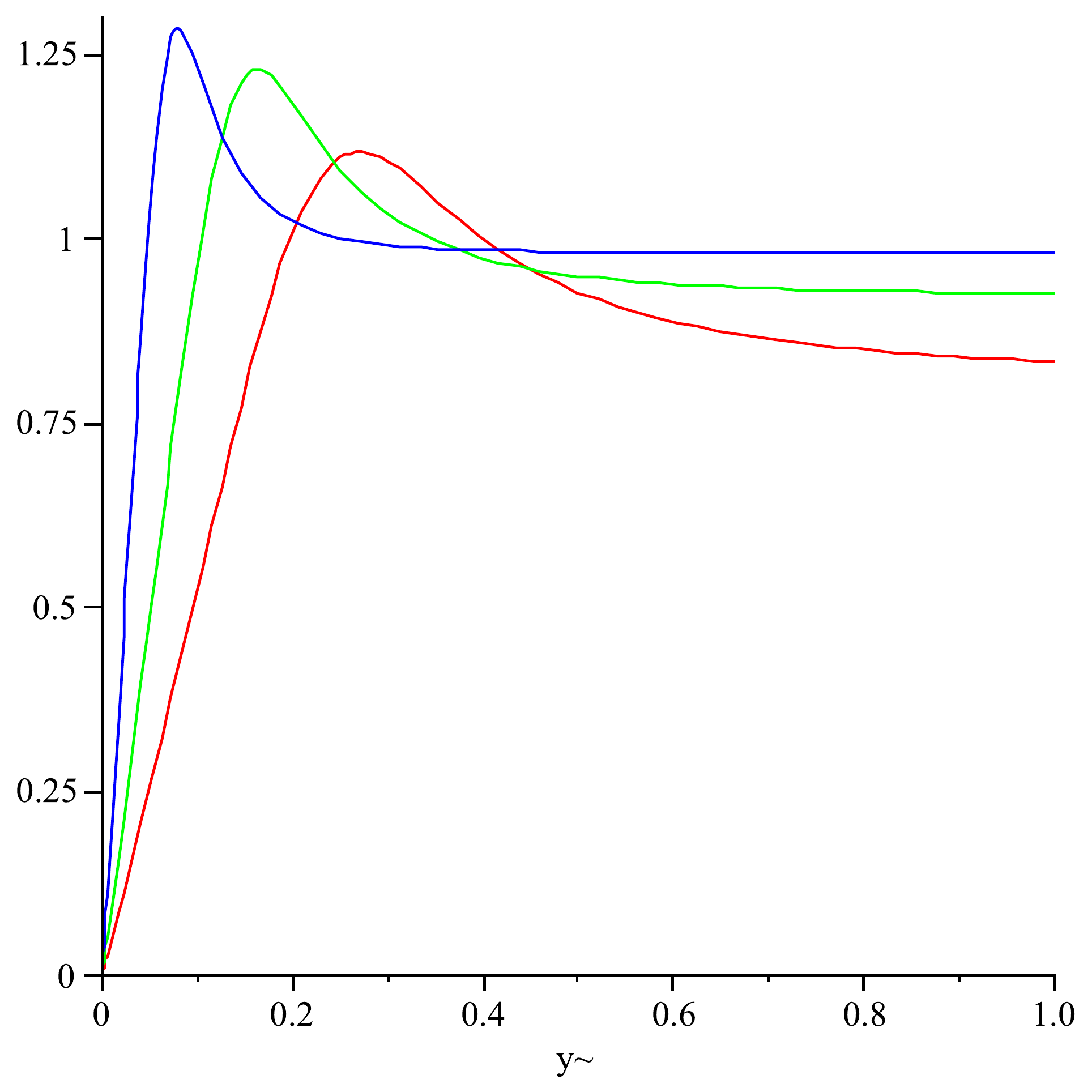}
                    \caption{(Left) The density of complex zeros of a random $SO(2)$ polynomial for $N=10,25,100$.  Because of symmetries, it is sufficient to plot the density along the imaginary axis for $0<y\leq 1.$ Here we have normalized so that the density of zeros of a random $SU(2)$ polynomial is the constant function 1. (Right) The density of complex critical points of a random $SO(2)$ polynomial for $N=10,25,100$, plotted along the imaginary axis for $0< y\leq 1.$  Again, we have normalized so that the density of critical points of a random $SU(2)$ polynomial is the constant function 1.  In both cases, the density is converging to 1.}\label{zeros1}
            \end{figure}

     The density of real (resp. complex) zeros of random polynomials in one and several variables with real (resp. complex) Gaussian coefficients has been studied by many.  See, for example, \cite{kac}, \cite{rice}, \cite{bbl92}, \cite{bbl96}, \cite{hannay}, and \cite{ek}. In one variable, Shepp and Vanderbei \cite{sv95}, Ibragimov and Zeitouni \cite{zeitouni}, and Prosen \cite{prosen} have studied \textit{complex} zeros of \textit{real} polynomials.  Prosen followed Hannay's approach and found both an unscaled and a scaled density formula for the complex zeros of a random polynomial with independent real Gaussian coefficients.  One consequence of Prosen's unscaled density formula is that, away from the real line, the density of complex zeros of a random $SO(2)$ polynomial, which is the polynomial given by
       $$ f_N(z) = \sum_{j=0}^N a_j {N \choose j}^{\frac{1}{2}} z^j, $$
    where $a_j$ is a \textit{real} standard Gaussian random variable, rapidly approaches the density of complex zeros of a random $SU(2)$ polynomial, which is the polynomial given by
       $$ f_N(z) = \sum_{j=0}^N c_j {N \choose j}^{\frac{1}{2}} z^j, $$
    where $c_j$ is a \textit{complex} standard Gaussian random variable), as the degree of the polynomial goes to infinity.  In \cite{bmzeros}, the author used the Poincar\'{e}-Lelong formula to show this convergence, recovering Prosen's single variable result \cite{prosen} for the $SO(2)$ polynomials, and also showed the convergence to be exponential.  In Theorem 1 in \cite{bmzeros}, the author generalized this result to the density of zeros of a random $SO(m+1)$ polynomial system in $m$ variables (defined below).  Figure \ref{zeros1} illustrates this convergence in the case $m=1$.  Note that the density functions are normalized so that the density in the complex coefficients case is the constant function 1.  In this paper, we use a generalized Kac-Rice formula for systems of real polynomials to prove an analogous result for the density of critical points of a random $SO(m+1)$ polynomial in $m$ variables.  This convergence can also been seen in Figure \ref{zeros1}.

    \subsection{Density of zeros}
    Consider $\mathbf{h} _{m,N}=(f_{1,N}, ...\, , f _{m,N}):\mathbb{C}^m \rightarrow \mathbb{C}^m $, where $f _{q,N}$ is a polynomial of the form
        \begin{align}
            f_{q,N}(\mathbf{z}) = \sum_{|J|=0}^{N} c_J^q {N\choose J} ^{1/2} \mathbf{z}^J,
        \end{align}
         where the $c^q_J$'s are independent complex random variables, where the random vector $\mathbf{c} = \{c^q_J\}\in \mathbb{C}^{D_N}, D_N = {N+m \choose m}$, has associated measure $d\gamma$, and where we are using standard multi-index notation.  Let
        \begin{align}\label{measures}
             d\gamma_{cx} &=
            \frac{1}{\pi^{N}}e^{-|\mathbf{c}|^2} d\mathbf{c},  \\
             \notag d\gamma_{real} &= \delta_{\mathbb{R}^{D_N}} \frac{1}{(2\pi)^{N/2}}e^{-|\mathbf{c}|^2/2} d\mathbf{c},
        \end{align}
         where $  \mathbf{c} \in \mathbb{C}^{D_N},$ and $\delta_{\mathbb{R}^{D_N}}$ is the delta measure on ${\mathbb{R}^{D_N}} \subset \mathbb{C}^{D_N}.$  Here $d\gamma_{cx}$ corresponds to the standard complex Gaussian coefficients case, where we are considering the random $SU(m+1)$ polynomial
        \begin{align}\label{fzeros}
            f_{q,N}(\mathbf{z}) = \sum_{|J|=0}^{N} c^q_J {N\choose J} ^{1/2} \mathbf{z}^J,
        \end{align}
        where the $c^q_J$'s are standard complex Gaussian random variables, and $d\gamma_{real}$ corresponds to the standard real Gaussian coefficients case, where we are considering the random $SO(m+1)$ polynomial
            $$ f^{}_{q,N}(\mathbf{z}) = \sum_{|J|=0}^{N} c^q_J {N\choose J} ^{1/2} \mathbf{z}^J = \sum_{|J|=0}^{N} a^q_J {N\choose J} ^{1/2} \mathbf{z}^J,$$
        where $c^q_J = a^q_J +i0$ is a standard real Gaussian random variable.

        Let $E_\gamma(\cdot)$ denote the expectation with respect to $\gamma$; or, in other words, integration over $\C^{D_N}$ with respect to the probability measure $d\gamma.$  Let
            $$\dis Z_{\mathbf{h}_{m,N}(\mathbf{z})} = \sum_{\mathbf{h}_{m,N}(\mathbf{z})=0} \kern -1em \delta_\mathbf{z}\,\,$$
        denote the distribution corresponding to the zeros of $\mathbf{h}_{m,N}(\mathbf{z}).$  Here, $\delta_\mathbf{z}$ is the Dirac delta function at $\mathbf{z}$, so $Z_{\mathbf{h}_{m,N}(\mathbf{z})}$ is a collection of deltas located at the zeros of $\mathbf{h}$.  $E_\gamma(Z_{\mathbf{h}_{m,N}(\mathbf{z})})$ denotes the density of the zeros of $\mathbf{h}$ with respect to the measure $d\gamma.$  We now restate the result in \cite{bmzeros} on the density of zeros:
        \begin{mainthm}[Theorem 1 in \cite{bmzeros}]\label{thm1}
                 \begin{align*}
                     \dis E_{\gamma_{real}}(Z_{\mathbf{h}_{m,N}(\mathbf{z})} ) &=  E_{\gamma_{cx}}(Z_{\mathbf{h}_{m,N}(\mathbf{z})} ) +O(e^{-\lambda_\mathbf{z} N}),
                \end{align*}
            for all $\mathbf{z} \in \mathbb{C}^m \backslash \mathbb{R}^m,$ where $\lambda_\mathbf{z}$ is a positive
            constant that depends continuously on $\mathbf{z}$.  The explicit formula for $\lambda_\mathbf{z}$ is
                \begin{align}\label{lambdazeros}
                   \lambda_\mathbf{z} = - \log\Big|\frac{1+\mathbf{z}\cdot \mathbf{z}}{1+||\mathbf{z}||^2}\Big|.
                \end{align}
            Also, for compact sets $K \subset \mathbb{C}^m\backslash\mathbb{R}^m$, the density converges uniformly with an error term of  $O(e^{-\lambda_K N})$, where $\lambda_K$ is a constant that depends only on $K$.
        \end{mainthm}

        Note that for $\mathbf{z} \in \mathbb{C}^m \backslash \mathbb{R}^m,$ the argument of the $\log$ is less than 1, and $\lambda_\mathbf{z}$ is positive.  The formula for $E_{\gamma_{cx}}(Z_{\mathbf{h}_{m,N}(\mathbf{z})} )$ is a special case of a result in \cite{ek}, and is a very simple function:
            \begin{align}
                E_{\gamma_{cx}}(Z_{\mathbf{h}_{m,N}(\mathbf{z})} ) = \frac{mN^m}{\pi^m}\frac{1}{(1+||\mathbf{z}||^2)^{m+1}}.
            \end{align}
        The formula for $E_{\gamma_{real}}(Z_{\mathbf{h}_{m,N}(\mathbf{z})} ) $ is very complicated, but, by this theorem, we know that $E_{\gamma_{real}}(Z_{\mathbf{h}_{m,N}(\mathbf{z})} ) $ equals a very simple function, $E_{\gamma_{cx}}(Z_{\mathbf{h}_{m,N}(\mathbf{z})} )$, plus some exponentially small term.

        Shiffman and Zelditch \cite{SZdist} and Bleher, Shiffman, and Zelditch (\cite{bszpl}, and \cite{bszuniv}) have generalized many results about random polynomials on $\mathbb{C}^m$ and $\mathbb{R}^m$ to real and complex manifolds.  In particular, in \cite{bszpl}, the authors use the Poincar\'{e}-Lelong formula to find a formula for the density of zeros and correlations between zeros.  In \cite{bmzeros}, we used this same approach to prove Theorem \ref{thm1}.
    \subsection{Density of Critical Points}
        In \cite{dszcrit}, Douglas, Shiffman, and Zelditch study the critical points of a holomorphic section of a line bundle over a complex manifold, motivated by applications in string theory.  They use a generalized Kac-Rice formula to find statistics of these complex critical points, namely the density of critical points and correlations between critical points.  In this paper, we study complex critical points of a random polynomial with \textit{real} coefficients and generalize the result in Theorem \ref{thm1} of \cite{bmzeros} to the density of critical points of a random $SO(m+1)$ polynomial. More precisely, let
            \begin{align}\label{complexpoly}
                \mathbf{h}_{m,N}(\mathbf{z}) = \sum_{|J|=0}^{N} c_J {N\choose J} ^{1/2} \mathbf{z}^J, \quad \mathbf{z} \in \mathbb{C}^m,
            \end{align}
            where the $c_J$'s are independent complex random variables, where the random vector $\{c_J\}\in \mathbb{C}^{D_N}, D_N = {N+m \choose m}$, has associated measure $d\gamma$, and where we are using standard multi-index notation.  Let $d\gamma_{cx}$ and $d\gamma_{real}$ be as defined in (\ref{measures}), and let
            $$C_{\mathbf{h}_{m,N}(\mathbf{z})} = \sum_{\mathbf{h}_{m,N}'(\mathbf{z})=0} \delta_\mathbf{z}$$
        be the critical points of $\mathbf{h}$. We prove the following:
            \begin{mainthm}\label{thmcrit} \addcontentsline{toc}{subsection}{Theorem \ref{thmcrit}}
                 We have
                    \begin{align*}
                        \dis E_{\gamma_{real}}(C_{\mathbf{h}_{m,N}(\mathbf{z})} ) &=  E_{\gamma_{cx}}(C_{\mathbf{h}_{m,N}(\mathbf{z})} ) +O(e^{-\lambda_\mathbf{z} N}),
                    \end{align*}
                for all $\mathbf{z} \in \mathbb{C}^m \backslash \mathbb{R}^m,$ where $\lambda_\mathbf{z}$ is a positive
                constant depending continuously on $\mathbf{z}$.  The explicit formula for $\lambda_\mathbf{z}$ is
                    \begin{align}\label{lambdacrit}
                        \lambda_\mathbf{z} = - \log\Big|\frac{1+\mathbf{z}\cdot \mathbf{z}}{1+||\mathbf{z}||^2}\Big|.
                    \end{align}
                Also, on compact sets $K \subset \mathbb{C}^m\backslash\mathbb{R}^m$, the convergence is uniform with an error term of  $O(e^{-\lambda_K N})$, where $\lambda_K$ is a constant that depends only on $K$.
            \end{mainthm}
         In other words, at any point away from $\mathbb{R}^m$, the expected density of critical points in the real coefficients case rapidly approaches the expected density of critical points in the complex coefficients case as $N$ gets large.  Note that $\lambda_\mathbf{z}$ in (\ref{lambdacrit}) and (\ref{lambdazeros}) are the same.

        Finding the density of critical points of $\mathbf{h}$ is equivalent to finding the density of simultaneous zeros of the $m$ partial derivatives of $\mathbf{h}$, or in other words, the density of zeros of $(f_1, \ldots, f_m): \C^m \rightarrow \C^m,$ where
            \begin{align}\label{fcrit}
                f_{q,N}(\mathbf{z}) = \frac{\partial h}{\partial z_q} = \sum_{|J|=0}^{N} c_J {N\choose J} ^{1/2} \frac{\partial}{\partial z_q}\mathbf{z}^J
            \end{align}
        Comparing (\ref{fcrit}) with (\ref{fzeros}), it seems at first glance that the critical points case in Theorem \ref{thmcrit} is very similar to the zeros case in Theorem \ref{thm1}.  However, it is more difficult than that case since the $m$ partial derivatives are not independent random functions.  The coefficients in (\ref{fcrit}) are the same for all $q$, while the coefficients in (\ref{fzeros}) are different for all $q$ (and independent).  This fact makes the Poincar\'{e}-Lelong method used in \cite{bszpl} and \cite{bmzeros} more difficult to apply when $m\geq 2$.   We instead use a generalized Kac-Rice formula for real systems similar to that used in \cite{dszcrit}.  If $m=1$, ${h}$ is a polynomial in one variable and has just one partial derivative, so there is no problem with dependent partial derivatives and we can follow the Poincar\'{e}-Lelong method in \cite{bmzeros}.

    \subsection{An exact formula in one variable}
        In one variable, both $ E_{\gamma_{real}}(C_{{h}_{m,N}({z})} )$ and $E_{\gamma_{cx}}(C_{{h}_{m,N}({z})} )$ are simple enough to write down.  We consider the polynomial
            $$\dis {h}_N({z}) = \sum_{\ell=0}^{N} c_\ell {N\choose \ell} ^{1/2} {z}^\ell,$$
        where $z\in\C$, where the $c_j$'s are independent complex random variables, and where the complex random vector $(c_\ell) \in \C^N$ has associated measure $d\gamma$. Let $d\gamma_{real}$ and $d\gamma_{cx}$ be as defined in \eqref{measures}, with $m=1$.  The critical points of ${h}$ correspond to the zeros of
            $$f_N(z) = \frac{\partial h}{\partial z} =   \sum_{\ell=0}^{N} c_\ell {N\choose \ell} ^{1/2} \frac{\partial}{\partial z} z^\ell.$$
        Using the Poincar\'{e}-Lelong formula, we can show that
            \begin{align}\label{complexcrit1}
                 E_{\gamma_{cx}}(C_{f_N}(z)) %
                &= \frac{N}{\pi}\left(\frac{1}{(1+|z|^2)^2}- \frac{2}{N(1+|z|^2)^2} + \frac{1}{(1+N|z|^2)^2} \right), %
            \end{align}
        We can also write
            \begin{align}\label{realcrit1}
                E_{\gamma_{real}}(C_{f_N}(z)) &= E_{\gamma_{cx}}(C_{f_N}(z)) + \tilde{E}_{N}(z),
            \end{align}
        where $\tilde{E}_{N}(z)$ is some ``error term," and we can show that
            \begin{align}\label{errorcrit1}
                \tilde{E}_{N}(z)
                &=  \frac{1}{\pi}  \frac{\partial^2}{\partial z \partial \overline{z}} \log
                \left(1+\sqrt{1-\left|\frac{(N^2z^2+N)(1+z^2)^{N-2}}{(N^2|z|^2+N)(1+|z|^2)^{N-2}}\right|^2}\right) \\
                &= O(e^{-\lambda_zN}).\notag
            \end{align}
        The steps used to obtain \eqref{complexcrit1}, \eqref{realcrit1}, and \eqref{errorcrit1} are very similar to the steps used in Section 2 of \cite{bmzeros}, and we omit the details here.

    \subsection{A scaling limit formula in one variable}
        Consider the scaling limit of the density,
             \begin{align}\label{scalinglimit}
                K_{\gamma_{}}^{\infty}({z})  = \lim_{N\rightarrow\infty}\frac{1}{N} E_{\gamma_{}}(C_{{h}_N(\frac{{z}}{\sqrt{N}})} ),
             \end{align}
        Using \eqref{complexcrit1}, \eqref{realcrit1}, \eqref{errorcrit1}, and \eqref{scalinglimit}, we get
            \begin{align}\label{scaledcrit1}
            \nonumber
                 K^\infty_{\gamma_{cx}}(z)&=\lim_{N\rightarrow\infty} \frac{1}{N} E_{\gamma_{cx}}(C_{f_N(\frac{z}{\sqrt{N}})})
                    =  \frac{1}{\pi} \left(1 + \frac{1}{(1+|z|^2)^2}\right), \text{ and }\\%
                K^\infty_{\gamma_{real}}(z) &=  K^\infty_{\gamma_{cx}}(z) + \tilde{E}^\infty_{\gamma_{real}}(z), \text{ where } \\ \nonumber
                    \tilde{E}^\infty_{\gamma_{real}}(z) &= \lim_{N\rightarrow\infty} \frac{1}{N} \tilde{E}_{N}(\frac{z}{\sqrt{N}})
                    = \frac{1}{  \pi } \frac{\partial^2}{ \partial z \partial   \overline{z}}  \log \left(1+\sqrt{1-\left|\frac{(1+z^2)e^{z^2}}{(1+|z|^2)e^{|z|^2}}\right|^2}\right). %
            \end{align}
        The formulas \eqref{errorcrit1} and \eqref{scaledcrit1} are similar to the corresponding formulas in \cite{bmzeros}.  However, note that \eqref{errorcrit1} does not have the same symmetries as the unscaled density of zeros in \cite{bmzeros}.  Also, in \cite{bmzeros}, the author shows that the scaled density of zeros tends linearly towards the real line and depends only on $ y = \im {z}:$
            \begin{align*}
                K^\infty_{\gamma_{real}}(z) = \frac{1}{\pi} \,
                \frac{1-(4y^2+1)e^{-4y^2}}{(1-e^{-4y^2})^{3/2}} = \frac{1}{\pi}\, y +O(y^3),
            \end{align*}
        for $y$ near 0.  For critical points, we still have that $K^\infty_{\gamma_{real}}(z)$ tends linearly toward zero as we approach the real line, but because of the additional $\frac{1+z^2}{1+|z|^2}$ term in (\ref{scaledcrit1}), the scaled density of critical points is no longer a function of only $y = \im z$:
            \begin{align}
                K^\infty_{\gamma_{real}}(z) =  \frac{1}{\pi} \frac{x^6 + 3x^4+6x^2 + 6}{(2+2x^2+x^4)^\frac{3}{2}}\,\, y + O(y^3),
            \end{align}
        for $y$ near $0$.
            \begin{figure}[h!]\label{scaledcrit}
            \centering
                \includegraphics[width=.45\textwidth]{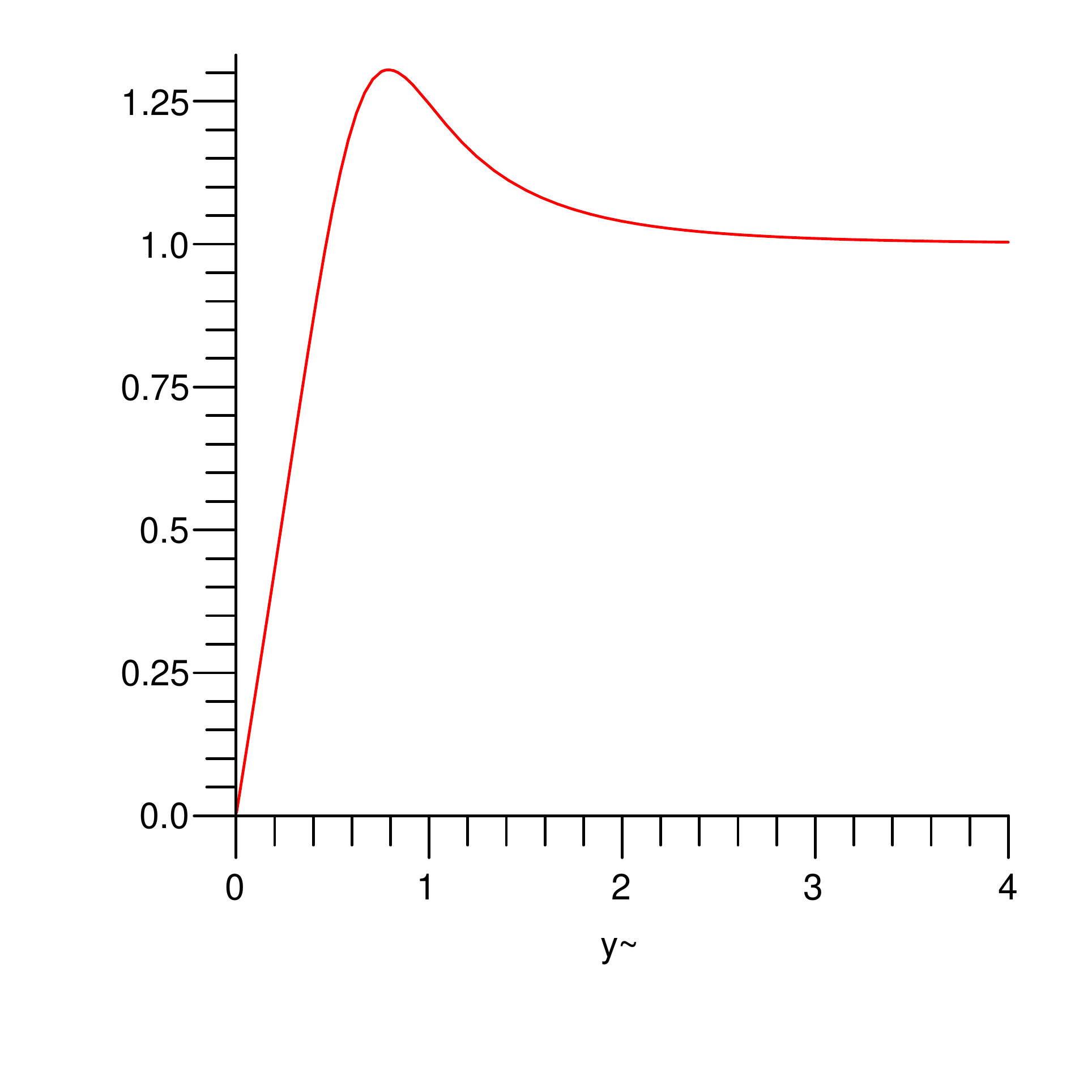}
                    \caption{The scaled density of critical points in one variable, plotted along the imaginary axis.}
            \end{figure}
        In Figure \ref{scaledcrit}, we plot (\ref{scaledcrit1}) along the imaginary axis, where we have the asymptotics
            \begin{align*}
                K^\infty_{\gamma_{real}}(z) =  \frac{3\sqrt{2} }{2\pi}\, y + O(y^3).
            \end{align*}

        The remainder of this paper is organized as follows.  In Section \ref{notation} we introduce some notation and state three intermediate results we will need to prove the main theorem.  In Sections \ref{proof1}-\ref{proof3} we prove these intermediate results. Finally, in Section \ref{pfthm}, we use the three results to prove Theorem \ref{thmcrit}.

\section{Some notation, and 3 intermediate results}\label{notation}
    We first consider a complex random polynomial $\mathbf{h}^{ }_{m,N}:\mathbb{C}^m \rightarrow
    \mathbb{C}$ of the form
        \begin{align*}
        \mathbf{h}^{}_{m,N}(\mathbf{z}) &= \sum_{|J|=0}^{N} c_{J}  {N\choose J} ^{1/2} \mathbf{z}^J,
        \end{align*}
    as described in \eqref{complexpoly}.  The following multi-index notation is being used:
        \begin{align*}
            \mathbf{z} &= (z_1,...,z_m),\\
            |J| &= j_1+ \cdots + j_m, \\
            c_J &= c_{j_1...j_m} \in \mathbb{C},\\
            {N\choose J} &=  {N\choose j_1, ..., j_m} = \frac{N!}{(N-j_1-...-j_m)!j_1!\,\,...\,\,j_m!}, \\
            \mathbf{z}^J &= z_1^{j_1} \cdots z_m^{j_m}.
        \end{align*}
    Instead of studying the critical points of this random
    polynomial $\mathbf{h}$, we could equivalently study the zeros of $(f^{ }_{1,N},
    ..., f^{ }_{m,N}):\mathbb{C}^m \rightarrow \mathbb{C}^m$, where
    $f^{}_{q,N}$ is a complex polynomial of the form
        \begin{align*}
            f^{}_{q,N}(\mathbf{z}) &=  \sum_{|J|=0}^{N} c_{J}  {N\choose J} ^{1/2}
            \frac{\partial }{\partial z_q} \mathbf{z}^J, \quad 1 \leq q \leq m.
        \end{align*}
    We'll consider $f^{}_{q,N}(\mathbf{z})$ as a function from
    $\mathbb{R}^{2m}$ to $\mathbb{R}^{2m},$ use the fact that
        $$C_{\mathbf{h}} =     Z_{f_1\cdots f_m}=Z_{f^r_1 \cdots f^r_m f^i_1 \cdots f^i_m},$$
    where $f_q = f_q^r + if_q^i, $ and find $E_\gamma(Z_{f^r_1 \cdots f^r_m
    f^i_1 \cdots f^i_m})$.

    Consider
        $$\mathbf{X}=(f_1^r,\ldots, f_m^r, {f}_1^i ,\ldots, {f}_m^i),$$
    let $z_q = x_q+iy_q$, and let ${\xi}$ be the matrix of derivatives of the function
        \begin{align*}
            (x_1,\ldots x_m, y_1 ,\ldots, y_m) \rightarrow {\mathbf{X}}
        \end{align*}
    from $\R^{2m} \rightarrow \R^{2m}$.
    We can write
        \begin{align*}
            {\xi} = \left(
                \begin{array}{cc} \dis%
                    \left(\frac{\partial f_q^r}{\partial x_{q'}}\right)_{1\leq
                    q,q'\leq m} & \dis\left(\frac{\partial f_q^r}{\partial
                    y_{q'}}\right)_{1\leq q,q'\leq m}
                    \\ \\%
                    \dis \left(\frac{\partial {f}_q^i}{\partial x_{q'}}\right)_{1\leq
                    q,q'\leq m} & \dis \left(\frac{\partial {f}_q^i}{\partial
                    y_{q'}}\right)_{1\leq q,q'\leq m}
                \end{array}
            \right)
        \end{align*}
    Note that since $\mathbf{h}$ and $f_q$ are holomorphic, the Cauchy-Riemann equations hold, giving
        \begin{align}\label{CRequations}
            \frac{\partial f^i_q}{\partial y_{q'}} &= \frac{\partial f^r_{q}}{\partial x_{q'}}, \quad
            \frac{\partial f^r_q}{\partial y_{q'}} = -\frac{\partial f^i_{q}}{\partial x_{q'}} .
        \end{align}
    Note also that since $\dis f^r_q = \frac{\partial \mathbf{h}^r}{ \partial x_q}$
                    and  $\dis f^i_q = \frac{\partial \mathbf{h}^i}{ \partial x_q}$,
    we have
        \begin{align}\label{mixedpart1}
            \frac{\partial f^r_q   }{\partial x_{q'}}   = \frac{\partial^2 \mathbf{h}^r}{\partial x_q' \partial x_q }
                                                        = \frac{\partial^2 \mathbf{h}^r}{\partial x_q  \partial x_q'} =
            \frac{\partial f^r_{q'}}{\partial x_{q }},
        \end{align}
    and likewise,
        \begin{align}\label{mixedpart2}
              \frac{\partial f^i_q}{\partial x_{q'}}
            = \frac{\partial f^i_{q'}}{\partial x_{q}}.
        \end{align}
     In light of \eqref{CRequations}, \eqref{mixedpart1}, and \eqref{mixedpart2}, we can choose a new basis and write
        \begin{align*}
            {\hat{{{\xi}}}} = [{\xi}]_\mathcal{{B}} = \left( \left(\frac{\partial
            f_q^r}{\partial x_{q'}}\right)_{q\leq q'}, \left({\frac{\partial
            f_q^i}{\partial x_{q'}}}\right)_{q\leq q'}\right) \in
            \mathbb{R}^{2d_m},
        \end{align*}
    where $d_m = m(m+1)/2.$  Below, $\xi$ will always refer to the matrix, $\hat{\xi}$ to the vector, and $\overline{\hat{\xi}}$ to the complex conjugate of the vector $\hat{\xi}$.

    Let ${\Delta}_{\gamma}$ be the covariance matrix of $\dis {{\mathbf{X}}
    \choose \hat{{\mathbf{\xi}}}}\in \R^{2m+2d_m}$ with respect to $\gamma$. We can write ${\Delta}_{\gamma}$ in block form as
        \begin{align}\label{bigdelta} \notag
            {\Delta}_{\gamma} &= \left(
                \begin{array}{cc}
                    {A}_{\gamma} & {B}_{\gamma} \\
                    {B}_{\gamma}^T & {C}_{\gamma}
                \end{array} \right) & \left((2m+2d_m) \times (2m+2d_m )\text{ matrix}\right)
            \\ \notag%
            {A}_{\gamma} &= \left(E_{\gamma}({x}_q
            \overline{x_{q'}})\right)_{q,q'} = {A}_{\gamma}^T%
            & (2m \times 2m \text{ matrix})
            \\
            {B}_{\gamma} &= \left(E_{\gamma}({x}_q
            \overline{\hat{{\mathbf{\xi}}}_j})\right)_{q,j} & (2m \times 2d_m \text{
            matrix})
            \\ \notag
            {C}_{\gamma} &= \left(E_{\gamma}(\hat{{\mathbf{\xi}}}_j
            \overline{\hat{{\mathbf{\xi}}}_{j'}})\right)_{j,j'} = {C}_{\gamma}^T &(2d_m \times 2d_m \text{
            matrix})
        \end{align}
    where $1 \leq {q},{q}' \leq 2m,$ and $  1 \leq {j},{j}' \leq 2d_m.$ Let ${\Lambda}_{\gamma}^{-1}$ be the lower right block of
    ${\Delta}^{-1}_{\gamma}$.  Then we have
        \begin{align}\label{lambdagamma}
            {\Lambda}_{\gamma} = {C}_{\gamma} - {B}^T_{\gamma}
            {A}^{-1}_{\gamma} {B}_{\gamma}.
        \end{align}

     We now state three results which will be used to prove the theorem.  In the first result, we reduce finding $E_{\gamma}(C_{\mathbf{h}})$ to finding a term we call $E_{{\Lambda}_{\gamma}}(\det{{\xi}}).$
        \begin{lemma}[Using the Kac-Rice formula]\label{kr}
            We have $\dis
            E_{\gamma}(C_{\mathbf{h}})
            = \frac{1}{\pi^{m} \sqrt{\det {A}_{\gamma}} } E_{{\Lambda}_{\gamma}}(\det
            {{\xi}}).
            $
        \end{lemma}
    \n Next, we use Wick's formula to write $E_{{\Lambda}_{\gamma}}(\det{{\xi}})$ in terms of entries of $\Lambda_\gamma.$
        \begin{lemma}[Using the Wick formula]\label{wick}
            $E_{{\Lambda}_{\gamma}}(\det {\xi}) = \phi({\Lambda_{\gamma}}),$ where
            $\phi({\Lambda_{\gamma}})$ is a homogeneous polynomial of degree $m$ in the entries of
            ${\Lambda_{\gamma}}.$
        \end{lemma}
    \n Finally, we show that ${{\Lambda}_{\gamma_{real}}}$ and ${{\Lambda}_{\gamma_{cx}}}$ differ by an exponentially small term.
        \begin{proposition}[A relationship between the real and complex Gaussian cases]\label{lambda}
                $$\dis\frac{{\Lambda}_{\gamma_{real}}}{(1+\mathbf{z}\cdot \overline{\mathbf{z}})^{N}} = \frac{{\Lambda}_{\gamma_{cx}}}{(1+\mathbf{z}\cdot \overline{\mathbf{z}})^{N}} + O(e^{-\lambda_\mathbf{z} N}), \mathbf{z} \in \mathbb{C}^m \backslash \mathbb{R}^m,$$
            where $\lambda_\mathbf{z}$ is given by the same formula as (\ref{lambdacrit}).
        \end{proposition}
    In the next three sections, we prove these results.  In the last section, we use these results to finish of the proof of Theorem \ref{thmcrit}.

\section{Proof of Lemma \ref{kr} - Using the Kac-Rice formula}\label{proof1}
     Note that because of Cauchy Riemann
    equations, $\det{\xi}$ is positive, and $\sqrt {\det \xi \xi ^T} =
    \det{\xi}$.  By the Kac-Rice formula for a system of real functions, we have
        \begin{align*}
            E_{\gamma}(Z_{f_1^r \cdots f_m^r {f}_1^i\cdots {f}_m^i}) =
            \int_{\mathbb{R}^{ 2d_m}} \sqrt{\det({\xi} {\xi}^T)}
            D_{\gamma}(0,\hat{{\mathbf{\xi}}};\mathbf{z}) d\hat{{\mathbf{\xi}}} = \int_{\mathbb{R}^{
            2d_m}} \det{\xi} \,\, D_{\gamma}(0,\hat{{\mathbf{\xi}}};\mathbf{z}) d\hat{{\mathbf{\xi}}}
        \end{align*}
    where $D_{\gamma}({\mathbf{X}},\hat{{\mathbf{\xi}}};\mathbf{z})$ is the Gaussian density in
    $2m + 2d_m$ real variables given by
        \begin{align*} \dis
            D_{\gamma}({\mathbf{X}},\hat{{\mathbf{\xi}}};\mathbf{z}) = \dis \frac{1}{\pi^{m+d_m}\sqrt{ \det
            {\Delta}_{\gamma}}} e^{ -\frac{1}{2}\langle {\Delta}_{\gamma}^{-1} {{\mathbf{X}}
            \choose \hat{\xi}}, {{\mathbf{X}} \choose \hat{{\mathbf{\xi}}}} \rangle}.
        \end{align*}
    Recall that in \eqref{bigdelta}, we wrote $\Delta_\gamma$ in block form.  Note that we can also write $A_\gamma,B_\gamma,$ and $C_\gamma$ in block
    form as
        \begin{align}\label{ABC}
            {A}_{\gamma} &= \left(
            \begin{array}{cc} \dis
                \left({E}_{\gamma}(f^r_{q}f^r_{q'})\right)_{q,q'} &
                \left({E}_{\gamma}(f^r_{q}f^i_{q'})\right)_{q,q'} \\ \\
                \left({E}_{\gamma}(f^i_{q}f^r_{q'})\right)_{q,q'} &
                \left({E}_{\gamma}(f^i_{q}f^i_{q'})\right)_{q,q'}
            \end{array} \right)
            \\ \\
            {B}_{\gamma} &=  \left(
            \begin{array}{cc}
                \left({E}_{\gamma}\left(f^r_{q}\frac{\partial f^r_{q'}}{\partial
                x_{p'}} \right)\right)_{q, \, q', p'} &
                \left({E}_{\gamma}\left(f^r_{q}\frac{\partial f^i_{q'}}{\partial
                x_{p'}} \right)\right)_{q, \, q', p'}  \\ \\
                \left({E}_{\gamma}\left(f^i_{q}\frac{\partial f^r_{q'}}{\partial
                x_{p'}} \right)\right)_{q, \, q', p'}  &
                \left({E}_{\gamma}\left(f^i_{q}\frac{\partial f^i_{q'}}{\partial
                x_{p'}} \right)\right)_{q, \, q', p'}
            \end{array} \right)
            \\ \\
            {C}_{\gamma} & =   \left(
            \begin{array}{cc}
                \left[{E}_{\gamma}\left(\frac{\partial f^r_{q}}{\partial
                x_p}\frac{\partial f^r_{q'}}{\partial x_{p'}} \right)\right]_{q,
                p, q', p'} & \left[{E}_{\gamma}\left(\frac{\partial
                f^r_{q}}{\partial x_p}\frac{\partial f^i_{q'}}{\partial x_{p'}}
                \right)\right]_{q,
                p, q', p'}  \\ \\
                \left[{E}_{\gamma}\left(\frac{\partial f^i_{q}}{\partial
                x_p}\frac{\partial f^r_{q'}}{\partial x_{p'}} \right)\right]_{q,
                p, q', p'}  & \left[{E}_{\gamma}\left(\frac{\partial
                f^i_{q}}{\partial x_p}\frac{\partial f^i_{q'}}{\partial x_{p'}}
                \right)\right]_{q, p, q', p'}
            \end{array} \right)
        \end{align}
    where $1 \leq {q}\leq p  \leq m,$ and $ 1\leq q'\leq p' \leq m.$

    Now, using the fact that for $D_{\gamma}(0,\hat{{\mathbf{\xi}}};\mathbf{z})$ only
    the lower right block of $\Delta_{\gamma}^{-1}$ matters, we can
    write
        \begin{align*} \dis
            D_{\gamma}(0,\hat{{\mathbf{\xi}}};\mathbf{z}) &=  \frac{1}{\pi^{m+d_m} \sqrt{\det
            {\Delta}_{\gamma}}} \exp\left( -\frac{1}{2}\langle {\Delta}_{\gamma}^{-1}
            \textstyle {0 \choose \hat\xi}, {0 \choose \hat\xi} \rangle\right)
            \\%
            &= \dis \frac{1}{\pi^{m} \sqrt{\det {A}_{\gamma}}} \frac{1}{\pi^{d_m}
            \sqrt{\det {\Lambda}_{\gamma}}} \exp \left( -\frac{1}{2}\langle
            {\Lambda}_{\gamma}^{-1} \hat{{\mathbf{\xi}}}, \hat{{\mathbf{\xi}}} \rangle \right) .
        \end{align*}

    \n We have also used the fact that $\, \det {\Delta}_{\gamma} = \det
    {A}_{\gamma} \det {\Lambda}_{\gamma}.$ We now have
        \begin{align*}
            E_{\gamma}(C_{\mathbf{h}}) &= \dis
            \frac{1}{\pi^{m} \sqrt{\det {A}_{\gamma}} }%
            \int_{\mathbb{R}^{d_m}} ({\det{\xi})} %
            \frac{1}{\pi^{d_m} \sqrt{\det {\Lambda}_{\gamma}}} \exp \left( -\frac{1}{2}\langle
            {\Lambda}_{\gamma}^{-1} \hat{{\mathbf{\xi}}}, \hat{{\mathbf{\xi}}} \rangle
            \right) d\hat{{\mathbf{\xi}}} \\ %
            &= \frac{1}{\pi^{m} \sqrt{\det {A}_{\gamma}} } E_{{\Lambda}_{\gamma}}(\det
            {{\xi}}).
        \end{align*}
    \endproof

\section{Proof of Lemma \ref{wick} - Using the Wick formula}\label{proof2}
    We now want to evaluate
    $E_{{\Lambda}_{\gamma}}(\det {\xi})$ using the Wick formula, which
    states that if $X_1, \ldots, X_{2m}$ are
    jointly Gaussian random variables, then %
        $$E(\prod_{q=1}^{2m} X_q ) = \sum \prod_{q=1}^{m} E(X_{i_q}X_{j_q})$$ %
    where the sum is over partitions of $\{1,...,2m\}$ into disjoint
    pairs $\{i_q, j_q\}$.  First we write
        \begin{align}\label{sumofproducts}
         E_{{\Lambda}_{\gamma}} (\det {\xi})  &= E_{{\Lambda}_{\gamma}} \left( \sum_{\sigma \in
        S_{2m}} sgn(\sigma) \prod_{q=1}^{2m} {\xi}_{q,\sigma(q)} \right) =
        \sum_{\sigma \in S_{2m}} sgn(\sigma) E_{{\Lambda}_{\gamma}} \left(
        \prod_{q=1}^{2m}
        {\xi}_{q,\sigma(q)} \right) \notag \\%
        &= \sum_{\sigma \in S_{2m}} sgn(\sigma) \sum \prod_{q=1}^{m}
        E_{{\Lambda}_{\gamma}} \left({\xi}_{i_q,\sigma(i_q)}
        {\xi}_{j_q,\sigma(j_q)} \right)
        \end{align}
    where $\sigma$ is a permutation, and where the second sum is over
    partitions of $\{1,...,2m\}$ into disjoint pairs $\{i_q, j_q\}$.  Note that terms of the form
        \begin{align*}
         E_{{\Lambda}_{\gamma}}\left({\xi}_{i_q,\sigma(i_q)}
            {\xi}_{j_q,\tau(j_q)}\right)
        \end{align*}
    are actually entries of ${\Lambda}_{\gamma}$.  So we have written
    $E_{{\Lambda}_{\gamma}}(\det {\xi})$ as a sum of products of
    $m$ entries in ${\Lambda}_{\gamma}.$ More specifically, we have that
        $$E_{{\Lambda}_{\gamma}}(\det {\xi}) = \phi({\Lambda_{\gamma}}),$$ where
    $\phi({\Lambda_{\gamma}})$ is a homogeneous polynomial in the entries of
    ${\Lambda_{\gamma}}.$

\section{Proof of Proposition \ref{lambda} - A relationship between the real and complex Gaussian cases}\label{proof3}
    Suppose now that we have the measures $d\gamma_{cx}$ and $d\gamma_{real} $ as defined in (\ref{measures}).
    Note that $d\gamma_{cx}$ corresponds to the standard complex Gaussian coefficients case, where we are considering
        $$ \mathbf{h}^{}_{m,N}(\mathbf{z}) = \sum_{|J|=0}^{N} c_{J}  {N\choose J} ^{1/2} \mathbf{z}^J $$
    where the $c_J$'s are standard complex Gaussian random variables, and $d\gamma_{real}$ corresponds to the standard real Gaussian coefficients case, where we are considering
        $$ \mathbf{h}^{}_{m,N}(\mathbf{z}) = \sum_{|J|=0}^{N} c_{J}  {N\choose J} ^{1/2} \mathbf{z}^J = \sum_{|J|=0}^{N} a_{J}  {N\choose J} ^{1/2} \mathbf{z}^J $$
    where $c_J = a_J +i0$ is a standard real Gaussian random variable.

    We now state four lemmas we need to prove the proposition.  Each result is a variation on the following theme:   on $\mathbb{C}^m \setminus \mathbb{R}^m,$ $\dis \frac{E_{\gamma_{cx}}(\cdot)}{(1+\mathbf{z}\cdot \overline{\mathbf{z}})^{N}}$ and  $\dis \frac{E_{\gamma_{real}}(\cdot)}{(1+\mathbf{z}\cdot \overline{\mathbf{z}})^{N}}$ are either equal, or differ by an exponentially small term $O(e^{-\lambda_\mathbf{z} N}).$
        \begin{lemma}\label{l3} Let $\mathbf{z}\in \mathbb{C}^m \setminus \mathbb{R}^m.$  The following are true for all $q,q',p,p'$:
        \begin{enumerate}
            \item $\dis \frac{E_{\gamma_{real}}(f_q f_{q'})}{(1+\mathbf{z}\cdot \overline{\mathbf{z}})^{N}}= \dis \frac{E_{\gamma_{cx}}  (f_q f_{q'})}{(1+\mathbf{z}\cdot \overline{\mathbf{z}})^{N}} + O(e^{-\lambda_\mathbf{z} N}) $
            \item $\dis \frac{{E}_{\gamma_{real}}\left(f_{q}\frac{\partial f_{q'}}{\partial z_{p'}} \right)}{(1+\mathbf{z}\cdot \overline{\mathbf{z}})^{N}}= \dis \frac{{E}_{\gamma_{cx}}\left(f_{q}\frac{\partial f_{q'}}{\partial
                z_{p'}} \right)}{(1+\mathbf{z}\cdot \overline{\mathbf{z}})^{N}} + O(e^{-\lambda_\mathbf{z} N})   $
            \item $\dis \frac{{E}_{\gamma_{real}}\left(\frac{\partial
                f_{q}}{\partial z_p}\frac{\partial f_{q'}}{\partial z_{p'}}
                \right)}{(1+\mathbf{z}\cdot \overline{\mathbf{z}})^{N}}= \dis \frac{{E}_{\gamma_{cx}}\left(\frac{\partial f_{q}}{\partial z_p}\frac{\partial f_{q'}}{\partial z_{p'}}
                \right) }{(1+\mathbf{z}\cdot \overline{\mathbf{z}})^{N}}+ O(e^{-\lambda_\mathbf{z} N}) $
        \end{enumerate}
        \end{lemma}
    \proof We prove just (1), and the rest are proved similarly.  For (1) we have
        \begin{align*}
        E_{\gamma_{cx}}  (f_q f_{q'}) %
        &= E_{\gamma_{cx}}  \left[\left(\sum_{|J|=0}^{N} c_{J}  {N\choose J} ^{1/2}
        \frac{\partial }{\partial z_q} \mathbf{z}^J \right) \left(  \sum_{|K|=0}^{N} c_{K}  {N\choose K} ^{1/2}
        \frac{\partial }{\partial z_{q'}} \mathbf{z}^K \right)\right] \\
        %
        &= \sum_{|J|=0}^{N}  \sum_{|K|=0}^{N}  E_{\gamma_{cx}}  \left( c_{J}c_{K}\right)  {N\choose J} ^{1/2}{N\choose K} ^{1/2}
        \frac{\partial }{\partial z_q} \mathbf{z}^J
        \frac{\partial }{\partial z_{q'}} \mathbf{z}^K = 0,
        \end{align*}
    since $E_{\gamma_{cx}}(c_Jc_K) = 0$ for all $J,K.$  Note that $E_{\gamma_{cx}} (c_J \overline{c_K}) = 1$ when $J=K$, so we have $E_{\gamma_{cx}}  (f_q \overline{f_{q'}}) \neq 0$; see Lemma \ref{l4}.
    Similarly, we have
        \begin{align*}
        \frac{E_{\gamma_{real}}  (f_q f_{q'})}{(1+\mathbf{z}\cdot \overline{\mathbf{z}})^{N}} %
        &=\frac{1}{(1+\mathbf{z}\cdot \overline{\mathbf{z}})^{N}}\sum_{|J|=0}^{N}  \sum_{|K|=0}^{N}  E_{\gamma_{real}}  \left( c_{J}c_{K}\right)  {N\choose J} ^{1/2}{N\choose K} ^{1/2}
        \frac{\partial }{\partial {z}_q} \mathbf{z}^J
        \frac{\partial }{\partial z_{q'}} \mathbf{z}^K   \\
        &=\frac{1}{(1+\mathbf{z}\cdot \overline{\mathbf{z}})^{N}}\sum_{|J|=0}^{N}  \sum_{|K|=0}^{N}  E_{\gamma_{real}}  \left( a_{J}a_{K}\right)  {N\choose J} ^{1/2}{N\choose K} ^{1/2}
        \frac{\partial }{\partial z_q} \mathbf{z}^J
        \frac{\partial }{\partial z_{q'}} \mathbf{z}^K   \\
        &=\frac{1}{(1+\mathbf{z}\cdot \overline{\mathbf{z}})^{N}}\sum_{|J|=0}^{N}   {N\choose J}
        \frac{\partial }{\partial z_q} \mathbf{z}^J
        \frac{\partial }{\partial z_{q'}} \mathbf{z}^J,
        \end{align*}
    since $ E_{\gamma_{real}}  \left( a_{J}a_{K}\right) =1, $ when $ J=K, $ and is zero otherwise.  We can then write
        \begin{align}\label{ztildetrick}
        \frac{E_{\gamma_{real}}  (f_q f_{q'})}{(1+\mathbf{z}\cdot \overline{\mathbf{z}})^{N}}
        &= \frac{1}{(1+\mathbf{z}\cdot \overline{\mathbf{z}})^{N}}\sum_{|J|=0}^{N}   {N\choose J}
        \frac{\partial }{\partial z_q} \mathbf{z}^J
        \frac{\partial }{\partial \tilde{z}_{q'}} \tilde{\mathbf{z}}^J\Big|_{\tilde{\mathbf{z}} = \mathbf{z}} \notag\\
        &= \frac{1}{(1+\mathbf{z}\cdot \overline{\mathbf{z}})^{N}}\frac{\partial }{\partial z_q} \frac{\partial }{\partial \tilde{z}_{q'}}\sum_{|J|=0}^{N}   {N\choose J}
        \mathbf{z}^J \tilde{\mathbf{z}}^J
         \Big|_{\tilde{\mathbf{z}} = \mathbf{z}} \\
        &= \frac{1}{(1+\mathbf{z}\cdot \overline{\mathbf{z}})^{N}}\frac{\partial }{\partial z_q} \frac{\partial }{\partial \tilde{z}_{q'}}(1+\mathbf{z}\cdot\tilde{\mathbf{z}})^N
         \Big|_{\tilde{\mathbf{z}} = \mathbf{z}} \notag\\
         &=\frac{N(N-1){z}_{q}{z}_{q'} }{(1+\mathbf{z}\cdot \overline{\mathbf{z}})^{2}}\left(\frac{1+\mathbf{z}\cdot{\mathbf{z}}}{1+\mathbf{z}\cdot \overline{\mathbf{z}}}\right)^{N-2}.\notag
        \end{align}
    Since $|1+\mathbf{z}\cdot \mathbf{z}| < |1+\mathbf{z} \cdot \overline{\mathbf{z}}| = 1+\mathbf{z} \cdot \overline{\mathbf{z}},$ for all $\mathbf{z}\in \mathbb{C}^m \setminus \mathbb{R}^m,$ we have that $\dis \Big| \frac{1+\mathbf{z}\cdot{\mathbf{z}}}{1+\mathbf{z}\cdot \overline{\mathbf{z}}}\Big| < 1$ for all $\mathbf{z}\in \mathbb{C}^m \setminus \mathbb{R}^m,$ which implies that $\dis \left(\frac{1+\mathbf{z}\cdot{\mathbf{z}}}{1+\mathbf{z}\cdot \overline{\mathbf{z}}}\right)^{N-2} = O(e^{-\lambda_\mathbf{z} N}), \mathbf{z}\in \mathbb{C}^m \setminus \mathbb{R}^m, $ where
        $ \dis    \lambda_\mathbf{z} = - \log\Big|\frac{1+\mathbf{z}\cdot \mathbf{z}}{1+||\mathbf{z}||^2}\Big|.$
    So we have
        $$ \frac{E_{\gamma_{real}}  (f_q f_{q'})}{(1+\mathbf{z}\cdot \overline{\mathbf{z}})^{N}}  = O(e^{-\lambda_\mathbf{z} N}),  \mathbf{z} \in\mathbb{C}^m \setminus \mathbb{R}^m.$$
    The results (2) and (3) in Lemma \ref{l3} can be proved similarly by defining $\tilde{z}$ and pulling the derivatives outside the sum as we did in \eqref{ztildetrick}.
    \endproof

    \begin{lemma}\label{l4}  
            We have  for all $q,q',p,p'$:
                \begin{enumerate}
                \item $\dis E_{\gamma_{real}} (f_q \overline{f_{q'}}) = E_{\gamma_{cx}}  (f_q \overline{f_{q'}})$
                \item $\dis {E}_{\gamma_{real}}\left(f_{q}\overline{\frac{\partial f_{q'}}{\partial
                z_{p'}}} \right) = {E}_{\gamma_{cx}}\left(f_{q}\overline{\frac{\partial f_{q'}}{\partial
                z_{p'}}} \right) $
                \item $\dis {E}_{\gamma_{real}}\left(\frac{\partial
                f_{q}}{\partial z_p} \overline{\frac{\partial f_{q'}}{\partial z_{p'}}}
                \right)= {E}_{\gamma_{cx}}\left(\frac{\partial
                f_{q}}{\partial z_p} \overline{\frac{\partial f_{q'}}{\partial z_{p'}}}
                \right)$
                \end{enumerate}
            %
            %
        \end{lemma}

    \proof We again prove just (1).  We have
        \begin{align*}
        E_{\gamma_{cx}}  (f_q \overline{f_{q'}}) %
        &= E_{\gamma_{cx}}  \left[\left( \sum_{|J|=0}^{N} c_{J}  {N\choose J} ^{1/2}
        \frac{\partial }{\partial z_q} \mathbf{z}^J \right) \left( \overline{\sum_{|K|=0}^{N} c_{K}  {N\choose K} ^{1/2}
        \frac{\partial }{\partial z_{q'}} \mathbf{z}^K }\right)\right] \\
        &= \sum_{|J|=0}^{N}  \sum_{|K|=0}^{N} E_{\gamma_{cx}}  (c_{J}\overline{c_{K}} ) {N\choose J} ^{1/2}{N\choose K} ^{1/2}
        \frac{\partial }{\partial z_q} \mathbf{z}^J
        \overline{\frac{\partial }{\partial z_{q'}} \mathbf{z}^K } \\
        &=\sum_{|J|=0}^{N}   {N\choose J}
        \frac{\partial }{\partial z_q} \mathbf{z}^J
        \overline{\frac{\partial }{\partial z_{q'}} \mathbf{z}^J},
        \end{align*}
    since $ E_{\gamma_{cx}}  \left( c_{J}\overline{c_{K}}\right) =1, $ when $ J=K, $ and is zero otherwise. Likewise, since $E_{\gamma_{real}}  (c_{J}\overline{c_{K}} ) = E_{\gamma_{real}}  (a_{J}\overline{a_{K}} ) = E_{\gamma_{real}}  (a_{J}{a_{K}} )$, and since $E_{\gamma_{real}}  (a_{J}{a_{K}} ) = 1$ when $J=K$ and is zero otherwise, we have
        \begin{align*}
        E_{\gamma_{real}}  (f_q \overline{f_{q'}}) %
        &=  \sum_{|J|=0}^{N}  \sum_{|K|=0}^{N} E_{\gamma_{real}}  (c_{J}\overline{c_{K}} ) {N\choose J} ^{1/2}{N\choose K} ^{1/2}
        \frac{\partial }{\partial z_q} \mathbf{z}^J
        \overline{\frac{\partial }{\partial z_{q'}} \mathbf{z}^K } \\
        &=\sum_{|J|=0}^{N}   {N\choose J}
        \frac{\partial }{\partial z_q} \mathbf{z}^J
        \overline{\frac{\partial }{\partial z_{q'}} \mathbf{z}^J} = E_{\gamma_{cx}}  (f_q \overline{f_{q'}}).
        \end{align*}
    By pulling the derivatives outside the sum as we did in \eqref{ztildetrick}, (2) and (3) can be proved similarly.
    \endproof

        \begin{lemma}\label{l5}  Let $\mathbf{z}\in\mathbb{C}^m \setminus \mathbb{R}^m.$  Using the results of the previous lemmas, we have, for all $q,q',p,p'$:
            \begin{enumerate}
                \item  $\dis \frac{E_{\gamma_{real}}(f^r_q f^i_{q'})}{(1+\mathbf{z}\cdot \overline{\mathbf{z}})^{N}} = \dis \frac{E_{\gamma_{cx}}  (f^r_q f^i_{q'}) }{(1+\mathbf{z}\cdot \overline{\mathbf{z}})^{N}}+  O(e^{-\lambda_\mathbf{z} N})$
                \item $\dis \frac{{E}_{\gamma_{real}}\left(f^r_{q}\frac{\partial f^i_{q'}}{\partial
                    x_{p'}} \right)}{(1+\mathbf{z}\cdot \overline{\mathbf{z}})^{N}} = \dis \frac{{E}_{\gamma_{cx}}\left(f^r_{q}\frac{\partial f^i_{q'}}{\partial
                    x_{p'}} \right) }{(1+\mathbf{z}\cdot \overline{\mathbf{z}})^{N}} + O(e^{-\lambda_\mathbf{z} N})   $
                \item $\dis \frac{{E}_{\gamma_{real}}\left(\frac{\partial
            f^r_{q}}{\partial x_p}\frac{\partial f^i_{q'}}{\partial x_{p'}}
            \right)}{(1+\mathbf{z}\cdot \overline{\mathbf{z}})^{N}}= \dis \frac{ {E}_{\gamma_{cx}}\left(\frac{\partial
                    f^r_{q}}{\partial x_p}\frac{\partial f^i_{q'}}{\partial x_{p'}}
                    \right)}{(1+\mathbf{z}\cdot \overline{\mathbf{z}})^{N}} + O(e^{-\lambda_\mathbf{z} N})   $
            \end{enumerate}
            Similar results hold for $f^r_q f^r_{q'}, f^i_q f^i_{q'}, $ and $f^i_q f^r_{q'}.$
        \end{lemma}
    \proof We again prove just (1).  Using $f^r_q = \frac{1}{2}(f_q+\overline{f_q}), f^i_q = \frac{1}{2i}(f_q-\overline{f_q}),$ and Lemmas \ref{l3} and \ref{l4}, we have
        \begin{align*}
        &\frac{E_{\gamma_{real}} (f^r_q f^i_{q'})}{(1+\mathbf{z}\cdot \overline{\mathbf{z}})^{N}}
        =  \frac{1}{4i}\left[
        \frac{E_{\gamma_{real}} (f_q f_{q'}) }                      {(1+\mathbf{z}\cdot \overline{\mathbf{z}})^{N}}
        -\frac{E_{\gamma_{real}} (f_q \overline{f_{q'}})}           {(1+\mathbf{z}\cdot \overline{\mathbf{z}})^{N}}
        +\frac{E_{\gamma_{real}} (\overline{f_q} f_{q'})}           {(1+\mathbf{z}\cdot \overline{\mathbf{z}})^{N}}
        -\frac{E_{\gamma_{real}} (\overline{f_q} \, \overline{f_{q'}})}{(1+\mathbf{z}\cdot \overline{\mathbf{z}})^{N}}\right] \\
        &= \frac{1}{4i} \left[
        \frac{E_{\gamma_{cx}} (f_q f_{q'})}{(1+\mathbf{z}\cdot \overline{\mathbf{z}})^{N}}
        + O(e^{-\lambda_\mathbf{z} N})
        -\frac{E_{\gamma_{cx}}(f_q \overline{f_{q'}})}{(1+\mathbf{z}\cdot \overline{\mathbf{z}})^{N}}
        +\frac{E_{\gamma_{cx}}(\overline{f_q} f_{q'})}{(1+\mathbf{z}\cdot \overline{\mathbf{z}})^{N}}
        -\frac{E_{\gamma_{cx}}(\overline{f_q} \, \overline{f_{q'}})}{(1+\mathbf{z}\cdot \overline{\mathbf{z}})^{N}}+ O(e^{-\lambda_\mathbf{z} N})
        \right]\\
        &=  \frac{E_{\gamma_{cx}} (f^r_q f^i_{q'})}{(1+\mathbf{z}\cdot \overline{\mathbf{z}})^{N}} + O(e^{-\lambda_\mathbf{z} N}).
        \end{align*}
    for all $\mathbf{z}\in \mathbb{C}^m \setminus \mathbb{R}^m.$  Statements (2) and (3) could be proved similarly, noting that $f_q$ is holomorphic and $\frac{\partial f_q}{\partial z_p} = \frac{\partial f_q}{\partial x_p}.$
    \endproof

    Using the Lemma \ref{l5}, we get the following results:
        \begin{lemma}\label{l6} We have
            \begin{enumerate}
                \item $\dis \frac{{A}_{\gamma_{real}}}{(1+\mathbf{z}\cdot \overline{\mathbf{z}})^{N}} = \frac{{A}_{\gamma_{cx}}}{(1+\mathbf{z}\cdot \overline{\mathbf{z}})^{N}} + O(e^{-\lambda_\mathbf{z} N})$
                \item $\dis\frac{{B}_{\gamma_{real}}}{(1+\mathbf{z}\cdot \overline{\mathbf{z}})^{N}} = \frac{{B}_{\gamma_{cx}}}{(1+\mathbf{z}\cdot \overline{\mathbf{z}})^{N}} + O(e^{-\lambda_\mathbf{z} N})$
                \item $\dis\frac{{C}_{\gamma_{real}}}{(1+\mathbf{z}\cdot \overline{\mathbf{z}})^{N}} = \frac{{C}_{\gamma_{cx}}}{(1+\mathbf{z}\cdot \overline{\mathbf{z}})^{N}} + O(e^{-\lambda_\mathbf{z} N})$
            \end{enumerate}
        \end{lemma}
    \proof
    \n From \ref{ABC} we have
        \begin{align}\label{Areal}
            &\frac{{A}_{\gamma_{real}}}{(1+\mathbf{z}\cdot \overline{\mathbf{z}})^{N}}
            = \left(
            \begin{array}{cc}
                \dis \left(\frac{{E}_{\gamma_{real}}(f^r_{q}f^r_{q'})}{(1+\mathbf{z}\cdot \overline{\mathbf{z}})^{N}}\right)_{q,q'} &
                \dis \left(\frac{{E}_{\gamma_{real}}(f^r_{q}f^i_{q'})}{(1+\mathbf{z}\cdot \overline{\mathbf{z}})^{N}}\right)_{q,q'}\\ \\
                \dis \left(\frac{{E}_{\gamma_{real}}(f^i_{q}f^r_{q'})}{(1+\mathbf{z}\cdot \overline{\mathbf{z}})^{N}}\right)_{q,q'} &
                \dis \left(\frac{{E}_{\gamma_{real}}(f^i_{q}f^i_{q'})}{(1+\mathbf{z}\cdot \overline{\mathbf{z}})^{N}}\right)_{q,q'}
            \end{array}
            \right),
        \end{align}
        and using Lemma \ref{l5}, we get that \eqref{Areal} equals
        \begin{align*}
            &= \left(
            \begin{array}{cc}
            \dis \left(\frac{{E}_{\gamma_{cx}}(f^r_{q}f^r_{q'})}{(1+\mathbf{z}\cdot \overline{\mathbf{z}})^{N}} +
                O(e^{-\lambda_\mathbf{z} N}) \right)_{q,q'}  &
            \dis \left(\frac{{E}_{\gamma_{cx}}(f^r_{q}f^i_{q'})}{(1+\mathbf{z}\cdot \overline{\mathbf{z}})^{N}} +
                O(e^{-\lambda_\mathbf{z} N}) \right)_{q,q'}  \\ \\
            \dis \left(\frac{{E}_{\gamma_{cx}}(f^i_{q}f^r_{q'})}{(1+\mathbf{z}\cdot \overline{\mathbf{z}})^{N}} +
                O(e^{-\lambda_\mathbf{z} N})\right)_{q,q'}  &
            \dis \left(\frac{{E}_{\gamma_{cx}}(f^i_{q}f^i_{q'})}{(1+\mathbf{z}\cdot \overline{\mathbf{z}})^{N}} +
                O(e^{-\lambda_\mathbf{z} N})\right)_{q,q'}
            \end{array} \right) \\
            &= \frac{{A}_{\gamma_{cx}}}{(1+\mathbf{z}\cdot \overline{\mathbf{z}})^{N}} + O(e^{-\lambda_\mathbf{z} N}),
       \end{align*}
    which proves (1).  The matrices $B_{\gamma_{real}}, B_{\gamma_{cx}},C_{\gamma_{real}},$ and $C_{\gamma_{cx}}$ can be written out similarly to get (2) and (3).
    \endproof
    Returning to the proof of the proposition, we can use \eqref{lambdagamma} and the Lemma \ref{l6} to get,
        \begin{align*}
            \frac{{\Lambda}_{\gamma_{real}}}{(1+\mathbf{z}\cdot \overline{\mathbf{z}})^{N}}
            & = \frac{{C}_{\gamma_{real}} - {B}^T_{\gamma_{real}}
            {A}^{-1}_{\gamma_{real}} {B}_{\gamma_{real}}}{(1+\mathbf{z}\cdot \overline{\mathbf{z}})^{N}}
            = \frac{{C}_{\gamma_{cx}} - {B}^T_{\gamma_{cx}} {A}^{-1}_{\gamma_{cx}} {B}_{\gamma_{cx}}}
                {(1+\mathbf{z}\cdot \overline{\mathbf{z}})^{N}}
                + O(e^{-\lambda_\mathbf{z} N})\\
            &=  \frac{{\Lambda}_{\gamma_{cx}}}{(1+\mathbf{z}\cdot \overline{\mathbf{z}})^{N}} + O(e^{-\lambda_\mathbf{z} N}).
        \end{align*}

\section{Proof of Theorem \ref{thmcrit}}\label{pfthm}
    By Lemma \ref{kr} we have
        \begin{align}\label{step1}
        E_{{\gamma_{real}}}(C_{\mathbf{h}_{N}(\mathbf{z})}) &= \frac{1}{\pi^m} \frac{E_{\Lambda_{\gamma_{real}}}(\det \xi)}{ \sqrt{\det
        A_{\gamma_{real}}}}
        = \frac{1}{\pi^m} \frac{1     }{  \sqrt{\frac{\det
        A_{\gamma_{real}}}{(1+\mathbf{z}\cdot \overline{\mathbf{z}})^{2Nm}} }} \frac{E_{\Lambda_{\gamma_{real}}}(\det \xi)}{ {(1+\mathbf{z}\cdot \overline{\mathbf{z}})^{Nm}}   }.
        \end{align}
    Note that from \eqref{sumofproducts} we can see that each term in the homogeneous polynomial $E_{\Lambda_{\gamma_{real}}}(\det \xi)=\phi(\Lambda_{\gamma_{real}})$ has $m$ factors, each of which is an element of $\Lambda_{\gamma_{real}},$ and likewise for $E_{\Lambda_{\gamma_{cx}}}(\det \xi)$.  This fact, along with Proposition \ref{lambda}, gives
        \begin{align}\label{elambda}
            \frac{{E}_{\Lambda_{\gamma_{real}}}(\det \xi)}{(1+\mathbf{z}\cdot \overline{\mathbf{z}})^{Nm}} = \frac{{E}_{\Lambda_{\gamma_{cx}}}(\det \xi)}{(1+\mathbf{z}\cdot \overline{\mathbf{z}})^{Nm}} + O(e^{-\lambda_\mathbf{z} N}),
         \end{align}
    and from \eqref{step1} and \eqref{elambda} we have
        \begin{align} \label{step2}
        E_{{\gamma_{real}}}(C_{\mathbf{h}_{N}(\mathbf{z})}) &= \frac{1}{\pi^m} \frac{1     }{  \sqrt{\frac{\det
        A_{\gamma_{real}}}{(1+\mathbf{z}\cdot \overline{\mathbf{z}})^{2Nm}} }} \left(\frac{E_{\Lambda_{\gamma_{cx}}}(\det \xi)}{ {(1+\mathbf{z}\cdot \overline{\mathbf{z}})^{Nm}}   }+O(e^{-\lambda_\mathbf{z} N})\right).%
        \end{align}
    Also note that $A_{\gamma}$ is a $2m \times 2m$ matrix, so each term in $\det A_{\gamma}$ has $2m$ factors, each of which is an element of $A_{\gamma}$.  Using this fact and Lemma \ref{l6}, we have
        \begin{align}\label{deta}
            \frac{\det {A}_{\gamma_{real}}}{(1+\mathbf{z}\cdot \overline{\mathbf{z}})^{2Nm}} = \frac{\det {A}_{\gamma_{cx}} }{(1+\mathbf{z}\cdot \overline{\mathbf{z}})^{2Nm}} + O(e^{-\lambda_\mathbf{z} N}),
        \end{align}
    and \eqref{step2} and \eqref{deta} give
        \begin{align}\label{step3}
        E_{{\gamma_{real}}}(C_{\mathbf{h}_{N}(\mathbf{z})})
        &= \frac{1}{\pi^m} \frac{1     }{  \sqrt{\frac{\det
        A_{\gamma_{cx}}}{(1+\mathbf{z}\cdot \overline{\mathbf{z}})^{2Nm}} + O(e^{-\lambda_\mathbf{z} N})}} \left(\frac{E_{\Lambda_{\gamma_{cx}}}(\det \xi)}{ {(1+\mathbf{z}\cdot \overline{\mathbf{z}})^{Nm}}   }+O(e^{-\lambda_\mathbf{z} N})\right)\\%
        \end{align}
    Simplifying \eqref{step3} further and using Lemma \ref{kr} again gives us
        \begin{align*}
        E_{{\gamma_{real}}}(C_{\mathbf{h}_{N}(\mathbf{z})})&= \frac{1}{\pi^m} \frac{1     }{  \sqrt{\frac{\det
        A_{\gamma_{cx}}}{(1+\mathbf{z}\cdot \overline{\mathbf{z}})^{2Nm}} }} \frac{E_{\Lambda_{\gamma_{cx}}}(\det \xi)}{ {(1+\mathbf{z}\cdot \overline{\mathbf{z}})^{Nm}}   }+O(e^{-\lambda_\mathbf{z} N})\\%
        &= \frac{1}{\pi^m} \frac{E_{\Lambda_{\gamma_{cx}}}(\det \xi)     }{  \sqrt{{\det
        A_{\gamma_{cx}}} }}   +O(e^{-\lambda_\mathbf{z} N})\\%
        &=E_{\gamma_{cx}}(C_{\mathbf{h}_{N}(\mathbf{z})}) + O(e^{-\lambda_\mathbf{z} N}).
        \end{align*}

\bibliographystyle{alpha}      
\bibliography{generalbib}   

\begin{thebibliography}{BSZ00b}

\bibitem[BBL92]{bbl92}
E.~Bogomolny, O.~Bohigas, and P.~Leboeuf.
\newblock {Distribution of roots of random polynomials}.
\newblock {\em Physical Review Letters}, 68(18):2726--2729, 1992.

\bibitem[BBL96]{bbl96}
E.~Bogomolny, O.~Bohigas, and P.~Leboeuf.
\newblock {Quantum chaotic dynamics and random polynomials}.
\newblock {\em Journal of Statistical Physics}, 85(5):639--679, 1996.

\bibitem[BSZ00a]{bszpl}
P.~Bleher, B.~Shiffman, and S.~Zelditch.
\newblock {Poincar{\'e}-Lelong Approach to Universality and Scaling of
  Correlations Between Zeros}.
\newblock {\em Communications in Mathematical Physics}, 208(3):771--785, 2000.

\bibitem[BSZ00b]{bszuniv}
P.~Bleher, B.~Shiffman, and S.~Zelditch.
\newblock {Universality and scaling of correlations between zeros on complex
  manifolds}.
\newblock {\em Inventiones Mathematicae}, 142(2):351--395, 2000.

\bibitem[DSZ04]{dszcrit}
M.R. Douglas, B.~Shiffman, and S.~Zelditch.
\newblock {Critical Points and Supersymmetric Vacua I}.
\newblock {\em Communications in Mathematical Physics}, 252(1):325--358, 2004.

\bibitem[EK95]{ek}
A.~Edelman and E.~Kostlan.
\newblock {How many zeros of a random polynomial are real?}
\newblock {\em American Mathematical Society}, 32(1):1--37, 1995.

\bibitem[Han96]{hannay}
J.H. Hannay.
\newblock {Chaotic analytic zero points: exact statistics for those of a random
  spin state}.
\newblock {\em J. Phys. A: Math. Gen}, 29(5):L101--L105, 1996.

\bibitem[IZ97]{zeitouni}
I.~Ibragimov and O.~Zeitouni.
\newblock {On Roots of Random Polynomials}.
\newblock {\em Transactions of the American Mathematical Society},
  349(6):2427--2441, 1997.

\bibitem[Kac48]{kac}
M.~Kac.
\newblock {On the Average Number of Real Roots of a Random Algebraic Equation
  (II)}.
\newblock {\em Proceedings of the London Mathematical Society}, 2(6):401, 1948.

\bibitem[Mac09]{bmzeros}
B~Macdonald.
\newblock {Density of Complex Zeros of a System of Real Random Polynomials}.
\newblock {\em Journal of Statistical Physics}, 136(5):807--833, 2009.

\bibitem[Pro96]{prosen}
T.~Prosen.
\newblock {Exact statistics of complex zeros for Gaussian random polynomials
  with real coefficients}.
\newblock {\em Journal of Physics A: Mathematical and General},
  29(15):4417--4423, 1996.

\bibitem[Ric54]{rice}
S.O. Rice.
\newblock {Mathematical analysis of random noise}.
\newblock {\em Selected Papers on Noise and Stochastic Processes}, pages
  133--294, 1954.

\bibitem[SV95]{sv95}
L.A. Shepp and R.J. Vanderbei.
\newblock {The Complex Zeros of Random Polynomials}.
\newblock {\em Transactions of the American Mathematical Society},
  347(11):4365--4384, 1995.

\bibitem[SZ99]{SZdist}
B.~Shiffman and S.~Zelditch.
\newblock {Distribution of Zeros of Random and Quantum Chaotic Sections of
  Positive Line Bundles}.
\newblock {\em Communications in Mathematical Physics}, 200(3):661--683, 1999.

\end{thebibliography}

\end{document}